\documentclass[letter]{ieice}
\usepackage[dvips]{graphicx}
\usepackage[fleqn]{amsmath}
\usepackage[varg]{txfonts}
\usepackage[psamsfonts]{amssymb}
\usepackage{balance}
\usepackage{cite}
\usepackage{latexsym}

\newtheorem{theorem}{Theorem} 
\newtheorem{definition}{Definition} 
\newtheorem{remark}{Remark}

\field{A}
\title[A Phenomenological Study on Threshold Improvement via Spatial Coupling]{A Phenomenological Study on Threshold Improvement via Spatial Coupling}
\titlenote{ 
The work of K.~Takeuchi was in part supported by the Grant-in-Aid for 
Young Scientists~(B) (No.~23760329) from MEXT, Japan.}
\authorlist{
 \authorentry[ktakeuchi@uec.ac.jp]{Keigo TAKEUCHI}{m}{UEC}
 \authorentry[tt@i.kyoto-u.ac.jp]{Toshiyuki TANAKA}{m}{KYOTO}
 \authorentry[kawabata@ice.uec.ac.jp]{Tsutomu KAWABATA}{m}{UEC}
}
\affiliate[UEC]{K.~Takeuchi and T.~Kawabata are with the Department of Communication Engineering and Informatics, the University of Electro-Communications, 
Tokyo 182-8585, Japan.}
\affiliate[KYOTO]{T.~Tanaka is with the Department of Systems Science, 
Graduate School of Informatics, Kyoto University, Kyoto 606-8501, Japan.}

\received{2011}{1}{1}
\revised{2011}{1}{1}

\begin{document}
\maketitle
\begin{summary}
Kudekar et al.\ proved an interesting result in low-density parity-check 
(LDPC) convolutional codes: The belief-propagation (BP) 
threshold is boosted to the maximum-a-posteriori (MAP) threshold by spatial 
coupling. Furthermore, the authors showed that the BP threshold for 
code-division multiple-access (CDMA) systems is improved up to the optimal 
one via spatial coupling. In this letter, a phenomenological model for 
elucidating the essence of these phenomenon, called threshold improvement, 
is proposed. The main result implies that threshold improvement occurs for  
spatially-coupled {\em general} graphical models.   
\end{summary}

\section{Introduction} \label{sec1} 
Low-density parity-check (LDPC) convolutional codes have been shown to 
outperform conventional LDPC block codes when iterative decoders 
based on belief propagation (BP) are used~\cite{Felstrom99,Sridharan04}. 
An LDPC convolutional code is constructed as a one-dimensionally 
coupled chain of LDPC block codes. 
As an explanation of this interesting result, it has been shown 
theoretically~\cite{Kudekar10,Kudekar11} and numerically~\cite{Lentmaier10} 
that the BP threshold of an LDPC convolutional code is  
boosted to the maximum-a-posteriori (MAP) one of the corresponding 
LDPC block code. This phenomenon is called ``threshold saturation'' via 
spatial coupling~\cite{Kudekar10,Kudekar11}. 

Recently, we showed that the threshold saturation also occurs in 
spatially-coupled code-division multiple-access (CDMA) systems: The 
BP threshold for sparsely-spread CDMA systems is boosted to the optimal one 
via spatial coupling~\cite{Takeuchi111,Takeuchi112}. Since it is unclear 
whether the BP threshold for {\em any} graphical model can be improved to 
the optimal one, the term ``threshold improvement'' via spatial coupling is 
used in this letter, instead of threshold saturation. 
It is believed that threshold improvement is universal, i.e., the 
performance of the BP algorithm can be improved by coupling general 
graphical models spatially. 
This conjecture has been verified for several graphical 
models~\cite{Hassani10,Kasai10}. The purpose of this letter is to present 
a phenomenological study that supports the universality of threshold 
improvement. 

Threshold improvement is a static property of spatially-coupled graphical 
models, rather than dynamical properties. 
It is well-known that BP can calculate the MAP solution exactly if the 
factor graph defining the BP is a tree~\cite{Pearl88}. 
Furthermore, if the BP algorithm for a general factor graph converges, 
the BP fixed-points correspond to 
the stationary points of the so-called Bethe free energy for the factor 
graph~\cite{Yedidia05}, while the MAP solution corresponds to the global 
minimizer of the true free energy. Roughly speaking, the Bethe free energy 
is obtained by locally approximating the original factor graph by trees. 
These results allow us to characterize 
the static properties of the BP algorithm by stationary solutions to a 
dynamical system that has a potential energy function whose fixed-points 
coincide with those for the Bethe free energy, which is analogous to the 
density evolution (DE) equation for LDPC codes. 
In this letter, we restrict graphical models to such a class of graphical 
models that BP algorithms converge asymptotically. 


As a phenomenological model for elucidating threshold improvement, we propose 
a spatially-coupled dynamical system with multiple stable solutions. 
This letter is organized as follows: In Section~\ref{sec2} 
the BP algorithm is characterized via a dynamical system with multiple stable 
solutions after presenting a motivating example for regular LDPC codes. 
In Section~\ref{sec3} a spatially-coupled dynamical system with multiple 
stable solutions is defined. Furthermore, an intuitive understanding of 
threshold improvement is presented. Section~\ref{sec4} presents the main 
result of this letter. 

\section{Systems without Spatial Coupling} \label{sec2}
\subsection{Density Evolution for regular LDPC Codes} 
The explicit formula of the Bethe free energy for LDPC codes is unknown. 
We shall construct potential energy associated with the Bethe 
free energy from DE. 
Let us consider the DE equation for regular LDPC codes 
over binary erasure channel (BEC) with erasure probability 
$\epsilon$~\cite[Theorem~3.50]{Richardson08}
\begin{equation} \label{DE} 
y_{t+1}-y_{t} = - \frac{dU}{dy}(y_{t};\epsilon), 
\quad y_{0}=1,  
\end{equation}
where the potential energy associated with the Bethe free energy is given by 
\begin{equation} \label{LDPC_potential} 
U(y;\epsilon) = \int_{0}^{y}\left\{
 z - \epsilon\lambda(c(z))  
\right\}dz,  
\end{equation}
with $c(y) = 1 - \rho(1-y)$. 
In these expressions, $y_{t}$ denotes the bit error rate (BER) for the 
BP decoder in iteration~$t$. Furthermore, $\lambda(y)$ and $\rho(y)$ are 
given by $\lambda(y)=y^{d_{\mathrm{v}}-1}$ and $\rho(y)=y^{d_{\mathrm{c}}-1}$, 
respectively, with $d_{\mathrm{v}}$ and $d_{\mathrm{c}}$ denoting the 
degrees of variable and check nodes. 
Note that the BP fixed-point $y_{\infty}$ corresponds to a stationary point 
of the potential energy~(\ref{LDPC_potential}). 
When $\epsilon$ is smaller than the BP threshold 
$\epsilon_{\mathrm{BP}}$, the potential~(\ref{LDPC_potential}) has the unique 
stable solution~$y_{\mathrm{l}}=0$. This observation implies that the BER 
converges to zero in $t\rightarrow\infty$ for 
$\epsilon<\epsilon_{\mathrm{BP}}$. When $\epsilon>\epsilon_{\mathrm{BP}}$, 
on the other hand, the potential~(\ref{LDPC_potential}) has two stable 
solutions $y_{\mathrm{l}}=0$ and $y_{\mathrm{r}}>0$, and one unstable solution 
$y_{\mathrm{u}}>0$, satisfying $y_{\mathrm{l}}<y_{\mathrm{u}}<y_{\mathrm{r}}$. 
The BER in this case converges to the strictly positive value~$y_{\mathrm{r}}$ 
in $t\rightarrow\infty$. Threshold saturation~\cite{Kudekar11} implies that 
$y_{t}$ can approach the smaller stable solution $y_{\mathrm{l}}=0$ for 
$\epsilon\in[\epsilon_{\mathrm{BP}},\epsilon_{\mathrm{MAP}})$ by spatial 
coupling, with $\epsilon_{\mathrm{MAP}}$ denoting the MAP threshold. 


\subsection{Dynamical System with Multiple Stable Solutions} 
\begin{figure}[t]
\begin{center}
\includegraphics{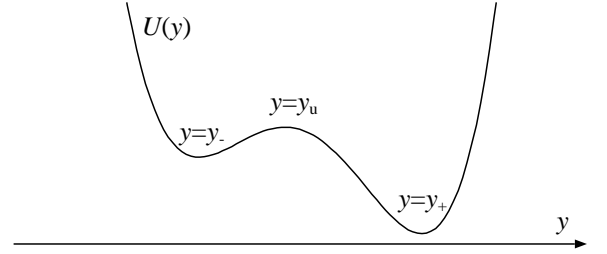}
\end{center}
\caption{
Bistable potential energy~$U(y)$. 
}
\label{fig1} 
\end{figure}

In order to investigate the universality of threshold improvement, 
we consider a continuous-time dynamical system with a {\em general} potential 
energy function $U(y)$ 
\begin{equation}
\frac{dy}{dt} = - \frac{dU}{dy}(y), 
\quad y(0) = y_{0}\in\mathbb{R}. 
\end{equation}
As a simple example, let us consider a bistable potential energy 
function~$U(y)$, shown in Fig.~\ref{fig1}. 
The potential energy $U(y)$ has two stable solutions and one unstable 
solution $y_{\mathrm{u}}$. Let $y_{-}$ and $y_{+}$ denote the smaller and 
larger stable solutions, respectively, i.e., $y_{-}<y_{+}$.  
If the initial value $y_{0}$ is larger than the unstable solution 
$y_{\mathrm{u}}$, the state~$y(t)$ converges to the larger stable solution 
$y_{+}$ in $t\rightarrow\infty$. Otherwise, $y(t)\rightarrow y_{-}$. 
Without loss of generality, we assume that larger $y(t)$ implies better 
performance. The larger stable solution $y_{+}$ corresponds to the optimal 
solution. Note that the term ``optimal solution'' is used to represent not 
the {\em global} stable solution but the {\em largest} stable solution in 
this letter. The typical BP solution corresponds to the smaller stable 
solution~$y_{-}$, because the initial value for the BP algorithm is commonly 
smaller than the unstable solution $y_{\mathrm{u}}$, e.g., 
see~\cite{Kabashima03,Takeuchi111} for CDMA systems. Thus, the inability of 
BP to get across the unstable solution makes the BP performance worse than the 
optimal performance when the potential energy $U(y)$ is not monostable. 

In this letter, the potential energy $U(y)$ is assumed to have multiple 
stable solutions. We hereafter refer to the largest stable 
solution, denoted by $y_{+}$, and the other stable solution(s) as 
the optimal solution and the BP solution(s), respectively. 

\section{Spatially-Coupled Dynamical System} \label{sec3} 
In order to obtain functionality of escaping from the BP solution(s), we 
consider a collection of identical systems coupled in a one-dimensional manner. 
Let $x\in(-x_{\max},\,x_{\max})$ denote the position of a system in the 
spatially coupling dimension, with $x_{\max}>0$ defining the size of the 
whole system, and let $y(x,t)$ be the state of the system at position $x$ and 
at time $t$. 
The spatially-coupled system we consider is governed by the equation 
\begin{equation} \label{system}
\frac{\partial y}{\partial t}
= -\frac{dU}{dy}(y) - \frac{1}{2}D'(y)\left(
 \frac{\partial y}{\partial x}
\right)^2
+ \frac{\partial}{\partial x}\left(
 D(y)\frac{\partial y}{\partial x}
\right),
\end{equation}
where $D(y)>0$ is a positive coupling function. 
We study the system~(\ref{system}) with the initial and boundary conditions 
\begin{equation}
y(x,0)=y_{0}\in\mathbb{R}
\quad \hbox{for $x\in(-x_{\max},x_{\max})$,} 
\end{equation}
\begin{equation}
y(\pm x_{\max},t) = y_{+}. 
\end{equation}

Let $(2N+1)$ denote the number of coupled systems. 
Threshold improvement occurs when the number of coupled systems, or 
equivalently $N$, tends 
to infinity and when the coupling strength~$\|D(y)\|$ tends to 
zero~\cite{Kudekar11,Takeuchi112}. 
The system~(\ref{system}) can be regarded as a space-continuum limit 
of ($2N+1)$ space-discrete coupled systems in $N\rightarrow\infty$, 
or an approximation of finite differences in the coupled systems. 
These interpretations are justifiable by letting $\Delta=x_{\max}/N$ and 
considering $(2N+1)$ copies of the original system at positions $x=i\Delta$ 
for $i=-N,-N+1,\ldots,N-1,N$. As $N$ gets large, the difference 
$y((i+1)\Delta,t)-y(i\Delta,t)$ is expected to be sufficiently small, so that 
one can approximate the difference by 
\begin{equation} \label{difference} 
y((i+1)\Delta,t)-y(i\Delta,t)\approx 
\Delta\frac{\partial y}{\partial x} 
+ \frac{\Delta^{2}}{2}\frac{\partial^{2}y}{\partial x^{2}}. 
\end{equation}
The difference~(\ref{difference}) is commonly included into a nonlinear 
function. Expanding the nonlinear function with respect to $\Delta$ yields 
a coupling function $D(y)$ depending on the state $y(x,t)$. 
See \cite{Takeuchi112} for spatially-coupled CDMA systems. 
The last two terms on the right-hand side of (\ref{system}) come from such 
approximations of space-discrete coupled systems. 

Threshold improvement for spatially-coupled CDMA systems can be understood 
from (\ref{system})~\cite{Takeuchi112}. Unfortunately, the phenomenological 
model~(\ref{system}) does not include LDPC convolutional codes. This is  
because a higher-order approximation is needed near the boundaries for an 
ensemble of LDPC convolutional codes~\cite{Kudekar11}, while the DE 
equation can be approximated by (\ref{system}) in the bulk region far from 
the boundaries. 

The system~(\ref{system}) can be regarded as 
a dynamical system with the so-called Ginzburg-Landau free 
energy functional~$H(y)$~\cite{Schmid66} as its potential energy
\begin{equation} \label{system1}
\frac{\partial y}{\partial t} = 
- \frac{\delta H}{\delta y}(y), 
\end{equation}
with 
\begin{equation} \label{potential} 
H(y) = \int_{-x_{\max}}^{x_{\max}}\left[
 U(y(x,t)) + \frac{D(y(x,t))}{2}\left(
  \frac{\partial y}{\partial x}  
 \right)^{2} 
\right]dx. 
\end{equation}
In (\ref{system1}), $\delta/\delta y$ denotes the functional derivative 
with respect to $y$. The energy functional~(\ref{potential}) implies that we 
impose  spatial coupling that smooths the state $y(x,t)$ spatially. 
The point of spatial coupling is that the boundaries are fixed 
to the optimal solution~$y_{+}$. A ``stretched rubber rope,'' both ends of 
which are fixed to the optimal solution~$y_{+}$, is utilized to ``climb'' the 
potential barriers between the BP solution(s) and the optimal solution. The 
tension of the rubber rope lifts the state~$y(x,t)$ toward the optimal 
solution. In LDPC convolutional codes, such a boundary condition results from 
termination of convolutional codes~\cite{Kudekar11}.


\section{Main Result} \label{sec4}
\begin{figure}[t]
\includegraphics[width=\hsize]{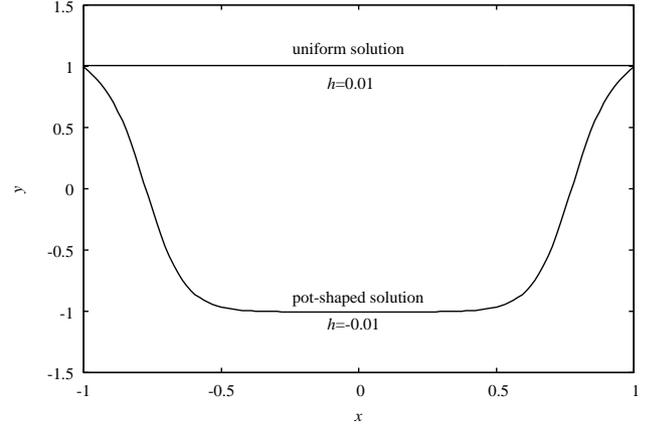}
\caption{
Stationary solutions to (\ref{system}) for $D=0.01$ and $x_{\max}=1$. 
The initial value $y_{0}$ is equal to $y_{-}$. 
}
\label{fig2}
\end{figure}

For simplicity, we hereafter assume that the coupling function does not 
depend on $y$, i.e., $D(y)=D>0$. We believe that the main result holds for 
state-dependent coupling functions.  
We focus on stationary solutions $y(x)$ to (\ref{system}), satisfying 
\begin{equation} \label{stationary_system} 
0 = - \frac{dU}{dy}(y) + D\frac{d^{2}y}{d x^{2}}, 
\quad y(\pm x_{\max}) = y_{+}. 
\end{equation} 
Figure~\ref{fig2} shows examples of the stationary 
solution~$y(x)$ for the double-well potential 
$U(y) = y^{4}/4 - y^{2}/2 - hy$ with the parameter $h\in\mathbb{R}$. 
The double-well potential has a metastable solution $y_{-}<0$ (resp.\ $y_{+}>0$) 
and a stable solution $y_{+}>0$ (resp.\ $y_{-}<0$) for $h>0$ (resp.\ $h<0$).  
When $h=0.01$, the state approaches the uniform solution 
$y(x)=y_{+}$. When $h=-0.01$, on the other hand, the stationary 
solution $y(x)$ is a pot-shaped solution. Note that this solution is a 
natural solution for the case where the state~$y(x,t)$ cannot climb potential 
barriers. Pot-shaped stationary solutions also appear in LDPC convolutional 
codes. See \cite{Kudekar11} for the details. 

The main result of this letter is that there are no pot-shaped stationary 
solutions if the boundaries are fixed to the global stable solution. 

\begin{definition}
A stationary solution $y(x)$ is called a pot-shaped solution if the 
following conditions are satisfied: 
\begin{itemize}
\item $y(0) < y(\pm x_{\max})$. 
\item $dy/dx\geq 0$ for $x>0$. 
\end{itemize}
\end{definition}

Note that any solution to (\ref{stationary_system}) is an even 
function of $x$, because the differential equation~(\ref{stationary_system}) 
is invariant under the transformation $x'=-x$. Thus, the second 
condition implies $dy/dx\leq0$ for $x<0$. 

\begin{theorem} \label{th1} 
Suppose that the coupling function $D(y)>0$ does not depend on $y$. 
If the largest stable solution $y_{+}$ of $U(y)$ is the unique global stable 
solution, then, there are no pot-shaped stationary solutions. 
\end{theorem} 

Before proving Theorem~\ref{th1}, we shall present the significance of 
Theorem~\ref{th1}. Obviously, the uniform solution $y(x)=y_{+}$ is a 
stationary solution to the spatially-coupled dynamical 
system~(\ref{system}). Intuitively, non-monotonic solutions on 
$x\in[0,x_{\max}]$ (or $x\in[-x_{\max},0]$) cannot become 
stationary solutions to (\ref{system}) with the uniform initial condition 
$y(x,0)=y_{0}$, because the state $y(x,t)$ should move closer to $y_{+}$ 
as the position $x$ gets closer to the boundary. Thus, 
Theorem~\ref{th1} implies that the state $y(x,t)$ converges to the uniform 
solution $y(x)=y_{+}$ in $t\rightarrow\infty$ for any initial value, i.e., 
the state can approach the optimal solution for general potential 
energy $U(y)$ such that the largest stable solution $y_{+}$ is the unique 
global stable solution. 

\begin{remark}
Suppose that the continuum approximation of the DE equation(s) for a 
spatially-coupled graphical model is given by (\ref{system}). 
We assume that the potential energy $U(y)$ is bistable and that it contains a 
parameter $\epsilon$, which corresponds to the erasure probability for LDPC 
codes over BEC or to the system load for CDMA systems. 
We rewrite the potential energy as $U(y;\epsilon)$. 
The potential $U(y_{-};\epsilon)$ at the smaller stable solution $y_{-}$ 
is assumed to be higher than that at the larger stable solution $y_{+}$ 
for small $\epsilon$, as shown in Fig.~\ref{fig1}. 
As $\epsilon$ increases, the potential $U(y_{-};\epsilon)$ at the smaller 
stable solution is assumed to get lower, while the potential at the larger 
stable solution gets higher. Eventually, the potential energy $U(y;\epsilon)$ 
may have the same height at the two stable solutions at 
$\epsilon=\epsilon_{\mathrm{BP}}^{(\mathrm{SC})}$. 
Theorem~\ref{th1} allows us to define the BP threshold for the spatially-coupled 
graphical model as the point $\epsilon_{\mathrm{BP}}^{(\mathrm{SC})}$ at 
which $U(y_{-};\epsilon_{\mathrm{BP}}^{(\mathrm{SC})})
=U(y_{+};\epsilon_{\mathrm{BP}}^{(\mathrm{SC})})$. 
The BP threshold $\epsilon_{\mathrm{BP}}^{(\mathrm{SC})}$ coincides  
with the optimal threshold for spatially-coupled CDMA 
systems~\cite{Takeuchi112}. It is unclear whether the BP threshold 
$\epsilon_{\mathrm{BP}}^{(\mathrm{SC})}$ coincides with the optimal one for 
{\em any} spatially-coupled graphical model. 
\end{remark}

{\em Proof of Theorem~\ref{th1}}: 
We shall prove Theorem~\ref{th1} by contradiction. 
Suppose that there is a pot-shaped stationary solution $y(x)$. 
Integrating (\ref{stationary_system}) after multiplying 
both sides by $dy/dx$, we obtain 
\begin{equation} \label{energy_conservation} 
\frac{D}{2}\left(
 \frac{dy}{dx}(x)
\right)^{2} = U(y) + C,  
\end{equation}
with a constant $C$. We use the boundary condition $y(\pm x_{\max})=y_{+}$ 
and the positivity of the left-hand side of (\ref{energy_conservation}) 
to find $C \geq - U(y_{+})$,   
where we have used the assumption that $y_{+}$ is the unique minimizer of 
$U(y)$. If $y_{+}$ was a local minimizer, the inequality would have 
to be replaced by $C\geq -U(y_{+}) + \Delta U$, with $\Delta U>0$ denoting 
the energy gap between $y_{+}$ and the global minimizer. 

Since we have assumed that there is a stationary solution satisfying 
$dy/dx\geq0$ (resp.\ $dy/dx\leq0$) for $x>0$ (resp.\ $x<0$),  
integrating (\ref{energy_conservation}) after taking the square root of both 
sides yields 
\begin{equation} \label{positive_solution} 
F(y) = x - \bar{x} \quad \hbox{for $x>0$,}
\end{equation}
with 
\begin{equation} \label{function_F} 
F(y) = \sqrt{\frac{D}{2}}\int_{\bar{y}}^{y}\frac{dy'}{\sqrt{U(y') + C}}, 
\end{equation}
where $\bar{y}$ denotes a value between $y(0)$ and $y_{+}$ that does not 
minimize the potential energy, i.e., $U(\bar{y})>U(y_{+})$. 
In (\ref{positive_solution}), we have selected a constant of integration 
such that $y(\bar{x})=\bar{y}$. 
Repeating the same argument for $x<0$, we obtain 
\begin{equation} \label{pot_shaped_solution}
y = F^{-1}(|x|-\bar{x}) \quad \hbox{for $x\in(-x_{\max},x_{\max})$,} 
\end{equation}
where $F^{-1}$ denotes the inverse function of (\ref{function_F}). 

Any stationary solution must be differentiable since it is a solution to 
the second-order differential equation~(\ref{stationary_system}). 
However, the solution~(\ref{pot_shaped_solution}) is not differentiable at 
the origin unless $dF^{-1}/dy|_{y=y(0)}=0$ or $dF/dy|_{y=y(0)}=\infty$, in 
which the value 
$y(0)$ at the origin is given by $y(0)=F^{-1}(-\bar{x})$. 
The uniqueness of the global stable solution implies that the integrand in 
(\ref{function_F}) can diverge only at $y=y_{+}$. Thus, 
a necessary condition for $dF/dy|_{y=y(0)}=\infty$ is 
$y(0)=y_{+}$, which contradicts $y(0)< y(\pm x_{\max})$. 
\QED 

\balance

\begin{figure}[t]
\includegraphics[width=\hsize]{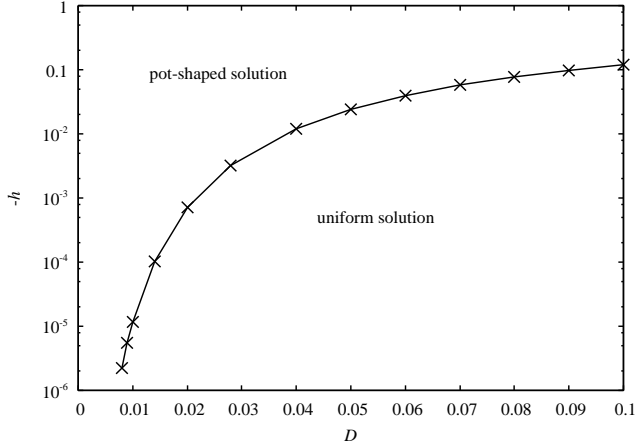}
\caption{
Bifurcation diagram for $x_{\max}=1$, obtained by numerical simulations of 
(\ref{system}). The initial value $y_{0}$ is equal to the smaller stable 
solution.   
}
\label{fig3}
\end{figure}

What occurs when the boundaries are fixed to a metastable solution?  
Figure~\ref{fig3} shows a bifurcation diagram of stationary solutions for the  
double-well potential $U(y) = y^{4}/4 - y^{2}/2 - hy$ with the parameter $h$. 
The larger stable solution $y_{+}$ for $h<0$ corresponds to the metastable 
solution, i.e., the boundaries are fixed to the metastable solution. 
The state converges to a pot-shaped stationary solution in $t\rightarrow\infty$ 
when $(D, -h)$ is located above the curve shown in Fig.~\ref{fig3}. Otherwise, 
the state converges to the uniform solution $y=y_{+}$. As long as $D$ is 
finite, the state may be conveyed to the uniform solution $y=y_{+}$ for small 
$h$. However, such a solution seems to disappear in $D\rightarrow0$, which 
corresponds to the limit in which threshold improvement 
occurs~\cite{Kudekar11,Takeuchi112}. 

\bibliographystyle{ieicetr}
\bibliography{kt-ieice2011_1}



\end{document}